          [2001/04/25 1.1 (PWD)]
\documentclass[a4paper,twocolumn]{esapub} 
\usepackage{natbib}
\usepackage{aastex_hack}
\usepackage{graphicx}

\title{Contribution of point sources to the soft $\gamma$-ray Galactic emission}
\author{R. Terrier}
\affil{Astroparticules et cosmologie, 11 place M. Berthelot, 75231 Paris, France}
\author{F. Lebrun}
\author{G. B\'elanger}
\author{A. Goldwurm}
\affil{Service d'Astrophysique, DAPNIA/CEA Saclay, 91911, Gif/Yvette, France}
\author{A. W. Strong}
\author{V. Schoenfelder}
\affil{MPE, Giessenbachstrasse, 85748 Garching, Germany}
\author{\\L. Bouchet}
\author{J.P. Roques}
\affil{CESR, 9 av. du Colonel Roche, BP 4346, 31028 Toulouse, France}
\author{A. Parmar}
\affil{ESA, RSSD, Keplerlaan 1, NL-2201, AZ Nordwijk, The Netherlands}

\begin{document}


\maketitle

\begin{abstract}
The nature of the soft gamma-ray (20-200 keV) Galactic emission has been a matter of debate for a long time. Previous experiments have tried to separate the point source contribution from the real interstellar emission, but with a rather poor spatial resolution, they concluded that the interstellar emission could be a large fraction of the total Galactic emission. INTEGRAL, having both high resolution and high sensitivity, is well suited to reassess more precisely this problem. Using the INTEGRAL core program Galactic Center Deep Exposure (GCDE), we estimate the contribution of detected point sources to the total Galactic flux.
\end{abstract}

\section{Introduction}

The interstellar emission is known to be mostly due to diffuse processes (that is interaction of cosmic rays with the interstellar medium) above 10 MeV \citep{Hunter97}. In the X-ray band, deep Chandra exposures have shown that up to 90\% of the total emission is due to purely diffuse emission \citep{Ebisawa01}. The situation is much less clear in the soft $\gamma$ ray domain: an intense and steep spectrum has been measured by OSSE up to 300 keV \citep{Kinzer99} where the hard positronium component starts to dominate (in the Galactic central regions). Explaining this spectrum has always been a very difficult task: $\gamma$-rays from 20 keV to 300 keV can be produced by Inverse Compton interaction of intermediate energy electrons (of the order of 100 MeV) \citep{Skibo96}. This is constrained by radio observations \citep{Webber83,Strong00}.
 
Another approach involves the bremsstrahlung of low energy electrons in the interstellar medium. But since the Coulomb losses are dominant under a few hundred MeV, most of the energy is dissipated through ionization. Therefore, the power required to sustain such a population of electron is very large (of the order of $10^{41} - 10^{43} erg s^{-1}$, that is the equivalent of the Glactic cosmic ray power). Besides, the large ionization should also produce an excessive dissociation of the molecules  in the central regions.
Hence, a major contribution from point sources was assumed \citep{Skibo96,Lebrun99}.

Observation of the diffuse emission in this energy range is a difficult task. OSSE with its $\mathrm{11^ox4^o}$ (FWHM) collimator had no imaging capabilities, and was only able to measure the total Galactic emission, i.e. point sources and diffuse emission, the resulting spectrum was therefore varying with time \citep{Kinzer99}. To remove the point source contribution, \citet{Purcell96} performed simultaneous observations with OSSE and SIGMA. The latter, with its coded mask system, had good imaging capabilities and a sensitivity of about 30 mCrab \citep{Paul91}. The total spectrum of sources detected by SIGMA was then subtracted from the total Galactic spectrum of OSSE. The resulting spectrum is of the order of  about 40 to 60\% of the total emission; the limited sensitivity of SIGMA suggests that a significant part of the residuals are due to unresolved point sources. Similar studies were made using OSSE and RXTE in the Scutum arm region (l=30°). The spectra of sources detected by RXTE, were extrapolated up to 50 keV and subtracted from the OSSE spectrum. A diffuse contribution of 50 to 70\% remained \citep{Valinia98}. The ballon-borne experiment HIREGS with its system of two collimators had the possbility to observe simultaenously small ($\le 3.7^o$) and large scale emission from the Galactic plane. By fitting the spectra of a few bright sources, \citet{Boggs00} found a diffuse emission spectrum much harder than previously.
One should also bear in mind, that observations of the Galactic ridge emission show non-thermal hard X-rays tails up to 20 keV. They have been detected in the Galactic plane \citep{Valinia00} as well as in the Galactic bulge \citep{Revnivtsev03}.

\begin{figure*}
\centering
\includegraphics[width=0.9\linewidth]{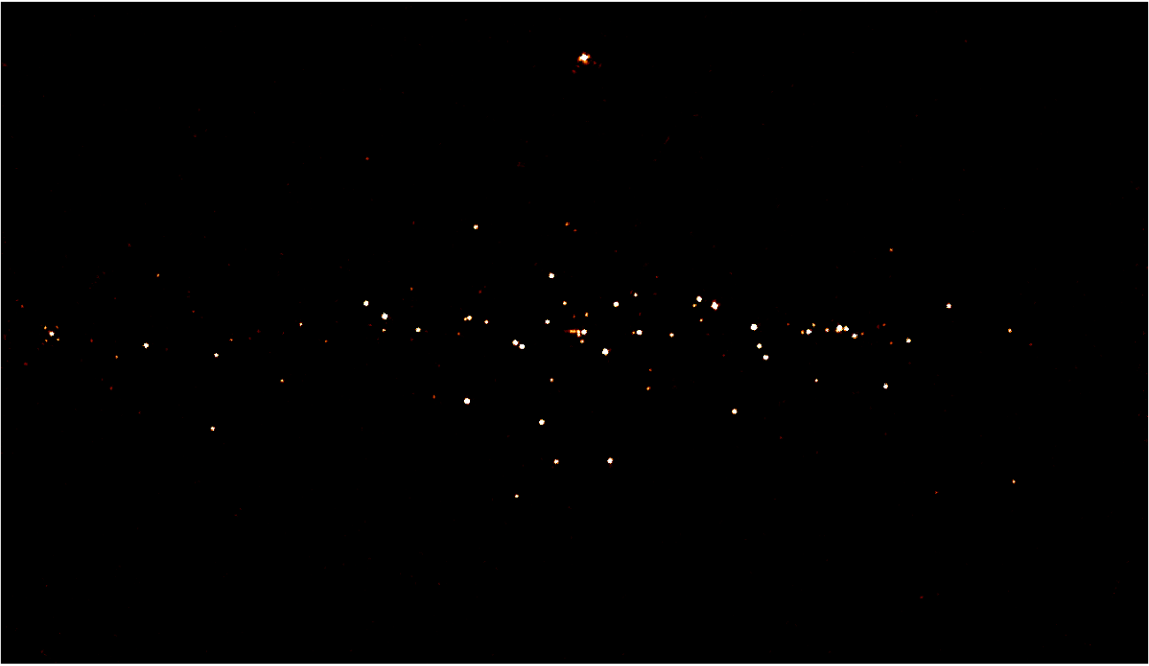}
\caption{Significance map of the GCDE survey. A total of 91 sources are detected above 6 $\sigma$.\label{fig:map}. Nearly half of them are identified as LMXB. This figure is taken from \citet{Lebrun04}}
\end{figure*}

Since one needs both a good angular resolution and a large field-of-view (FOV) to disentangle point sources from extended emission, IBIS \citep{Ubertini03}, the INTEGRAL imager,  with its low energy camera ISGRI \citep{Lebrun03}, seems well suited to perform this task. It has a good angular resolution of about 12' (FWHM), a wide FOV ($\mathrm{19^o}$ FWHM) and last but not least a mCrab sensitivity.  

\section{Compact sources extraction}

\subsection{The observations}

Observations of the whole Galactic Center Deep Exposure (GCDE) have been used as well as two Target of Opportunity observations in the Galactic central regions. This amounts to about 2000 science windows. They were analyzed with the standard analysis described in \citet{Goldwurm03}, in 4 energy bands 20-40 keV, 40-60 keV, 60-120 keV and 120-220 keV. In order to remove large scale residual background non-uniformities \citep{Terrier03}, the maps were filtered  and the variance was computed from local fluctutations in the intensity images. This variance is larger than the theoretical one and provides a more robust estimate of the significance and limits the possible false detections.

\subsection{Source detection}

Source detection was performed in each of the 20 - 40 , 20 - 60, 20 - 120 and 20 - 220 keV energy bands. Excesses above 6 $\sigma$ in any of these bands were considered and. 
A list of 91 sources candidates was produced, and counterparts were looked for in the SIMBAD database. About 40 Low Mass X ray Binaries (LMXB) were found, mostly in the Galactic bulge. 7 High-mass binaries (HMXB), a few plerionic supernovae remnants, pulsars, a few active Galactic nuclei and a soft $\gamma$-ray repeater are also detected. About 15 sources are identified but don't have known types (among which we find the new IGR sources), and about 15 have no acceptable counterparts in the databases.  Most of these results are in full agreement with the IBIS survey findings \citep{Bird04}.

\section{Using IBIS as a collimator}

The coded-mask imaging removes structures larger that the PSF, the diffuse emission is therefore totally washed out during the analysis. To compare it to the detected sources flux, one has to use IBIS as a collimator, that is compare point sources and total Galactic emission count rates. To do so, it is necessary to correct the measured count rates for the internal and cosmic background, and to estimate the source count rate using their flux estimated by the imaging.  

\subsection{Count rate corrections}

Here we assume that the measured count rate is due to:
\begin{itemize}
\item a time varying background caused by the cosmic-rays interactions
\item a contant and uniform background due to the diffuse cosmic and internal backgrounds
\item point sources
\item Galactic diffuse emission
\end{itemize}

\begin{figure}
\centering
\includegraphics[width=0.95\linewidth]{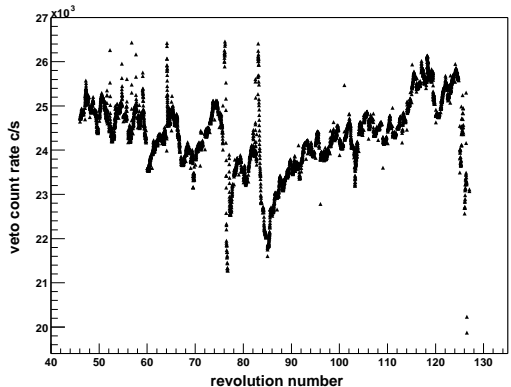}
\caption{IBIS veto count rate as a function of revolution number. There are strong variations of the cosmic-ray activity during the two GCDE periods which induce significant variations in the measured count rate. These variations are estimated using the ISGRI high energy count rate ($\geq$500 keV) which is strongly correlated to the veto count rate because it is mostly due to cosmic ray activation of the detector and its environment. \label{fig:veto}}
\end{figure}

The effects of the time variable background are very strong and have to be corrected: the variation of the veto count rate are large over the extent of the observations (see figure \ref{fig:veto}).
The high-energy count rate (E $\geq$ 500 keV) is mainly due to CR induced events: an on-axis source of 1 Crab would contribute at approximately 0.3 counts/s, which is more than two orders of magnitude lower than the average of 60 counts/s in this energy range. We will therefore use this information to estimate the level of time-variable background. The  correlation with low energy count rates is computed at high latitudes ($|b|\geq 30^o$) in regions depleted of point sources and Galactic emission, where all the count rate is due to cosmic ray induced and the isotropic  backgrounds. Using this correlation in the other pointings allows us to correct for the time variable background. The error in the coefficient determination introduces systematic errors in the corrected count rate.   

\begin{figure}
\centering
\includegraphics[width=0.9\linewidth]{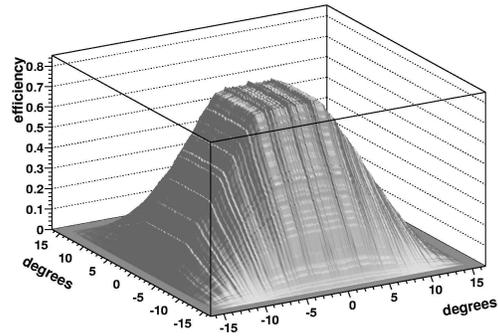}
\caption{Imaging efficiency applied to find the source count rate on the detector from its position in the FOV. It is the product of the acceptance (fraction of the detector pixels seen from a point in the sky) and the absorption.\label{fig:accept}}
\end{figure}

\subsection{Estimation of the source count rate on the detector}

The source fluxes estimated through the imaging are corrected for inclination on the telescope axis and averaged over all the observations. We therefore have to recover the sources count rate for individual science windows. To do this, we use the angular response of IBIS/ISGRI which is the product of the acceptance (the fraction of the detector actually light up through the mask) and the absorption and vignetting effects in the mask and mask structure. The resulting function is shown in figure \ref{fig:accept}. 

For each science window, the fluxes of all sources are estimated using their position in the FOV and then summed. Since there are some inaccuracies in the acceptance model, the count rates obtained from the image fluxes have to be calibrated on the Crab observed count rates. This calibration has been done in each energy band. Systematic errors of the order of 10\% are introduced this way.

\subsection{Looking for Galactic emission: fitting the residuals}

In order to improve the determination of the Galactic diffuse emission contribution, we assumed it is distributed uniformly along the Galactic plane with a $\mathrm{5^o}$ FWHM gaussian latitude distribution as suggested by \citet{Kinzer99}. To be more precise the CO and HI distribution could also be used \citep{Strong04}. The count rate due to interstellar emission is estimated by convolution of the acceptance response described above with this spatial distribution. The resulting latitude distribution of the diffuse emission count rates is approximately $\mathrm{15^o}$ wide (FWHM) because of the wide FOV. 

The corrected and source count rates of pointings lying between $\mathrm{l=-20^O}$ and $\mathrm{l=+20^O}$ are averaged in $2.5^o$ latitudes bins. The expected profile of the diffuse emission and a constant and isotropic background are then fitted to the residuals:
\[
\tau_{corr}(b)-\tau_{S}(b) = \alpha \tau_{diff}(b) + \beta 
\]

\section{Results}

\subsection{Longitude profiles}

\begin{figure}
\centering
\includegraphics[width=0.95\linewidth]{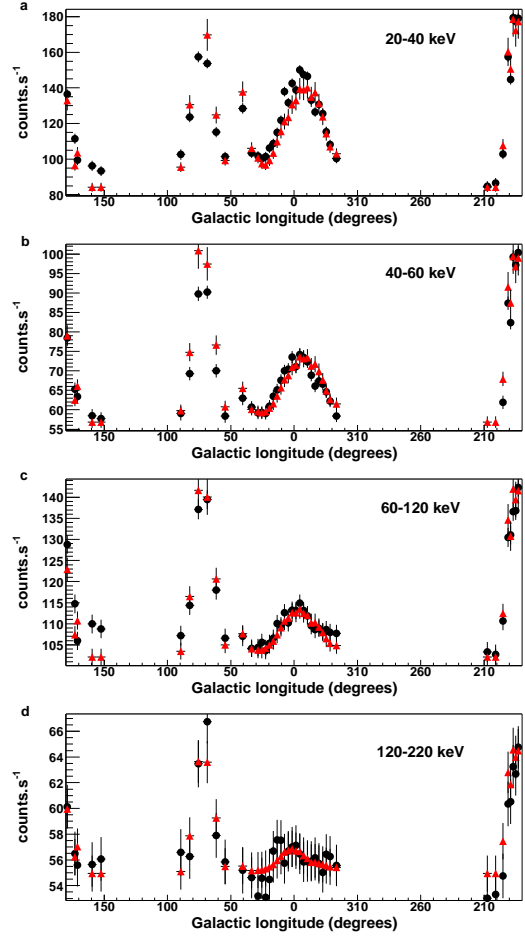}
\caption{Longitude profile of total Galactic emission. The black squares give the corrected count-rate, and the red triangles the sum of the estimated source count-rate and the constant isotropic background. Subfigures a to d are respectively in the 20 to 40, 40 to 60, 60 to 120 and 120 to 220 keV energy bands. This figure is taken from \citet{Lebrun04} \label{fig:lon}}
\end{figure}

In order to check for the consistency of this approach, the longitude profiles of corrected and source count rates have been compared. Only pointings close to the Galactic plane ($|b|<2^o$) have been considered. In each longitude bin, the mean count rate of all the science windows pointing in that bin have been computed. The constant background obtained through the latitude fit described above has been added to the source count rate in order to be comparable to the corrected count rate. The resulting profile resolution is limited by the $19^o$ FOV. The source and corrected count rate profiles are in good agreement in each of the 4 energy ranges, leaving only a small room to diffuse emission (see figure \ref{fig:lon}).

\begin{figure}
\centering
\includegraphics[width=0.95\linewidth]{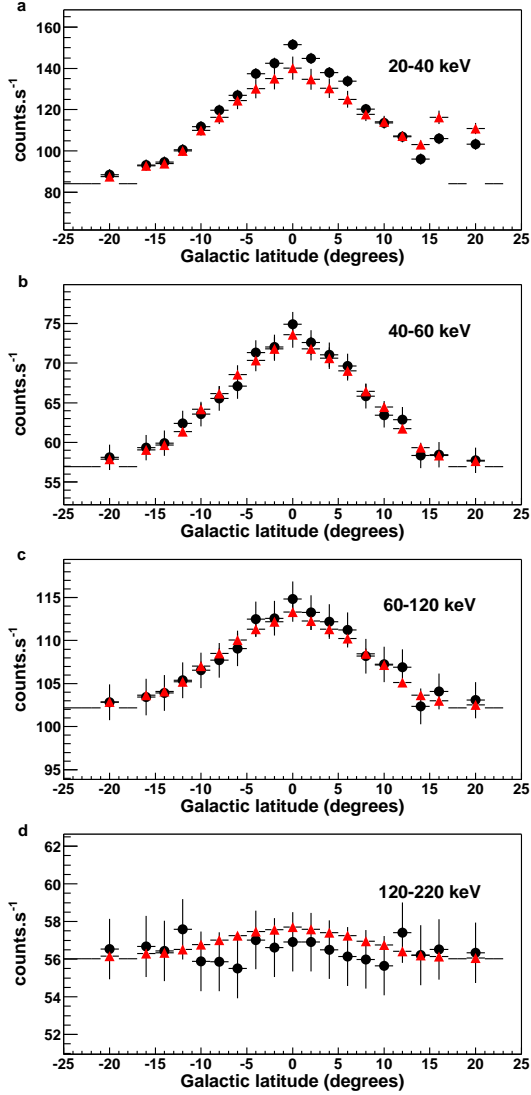}
\caption{Latitude profile of total Galactic emission. The black squares give the corrected count-rate, and the red triangles the sum of the estimated source count-rate and the constant isotropic background. Subfigures a to d are respectively in the 20 to 40, 40 to 60, 60 to 120 and 120 to 220 keV energy bands. A small fraction ($\sim$ 14\%) of the Galactic emission is not accounted for by point sources. At higher energies no significant residuals are detected.  In the last energy band, the transparency of shielding becomes large, which limits the relevance of the analysis. This figure is taken from \citet{Lebrun04} \label{fig:lat}}
\end{figure}

\subsection{Latitude profiles}

We show here the latitude profiles of the corrected and source count rates averaged in longitude from l=$340^o$ to l=$20^o$ (see figure \ref{fig:lat}). In the first energy band (20-40 keV), a small contribution of the order of 14\% is not accounted for by the detected sources. At higher energies, no significant residual emission is detected. The $3\sigma$ upper limit in the 40-60 keV band is 18\% and 27\% in the 60 to 120 keV energy band. At higher energies, the transparency of the shielding becomes large and the count rate correction is not any more valid. 
 The structure at $l=20^o$ in the low energy band is due to Sco-X1 whose flux is not well estimated because of remaining inaccuracies in the large angle off axis response of ISGRI. Since the systematic errors are still large a contribution less than 20\% of the total count-rate cannot be excluded.

\subsection{Diffuse emission spectrum}

\begin{figure}
\centering
\includegraphics[width=0.9\linewidth]{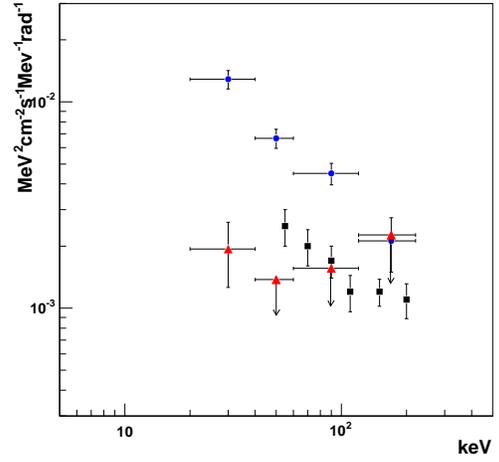}
\caption{Residual Galactic flux. The blue circles give the total source spectrum, the red triangles the remaining emission possibly of diffuse origin and the black squares give the estimated interstellar emission taking into account SIGMA sources \citep{Purcell96,Kinzer99}. The diffuse spectrum  is consistent with the one obtained by SPI with the same source list \citep{Strong04} \label{fig:spec}}
\end{figure}

The corrected count rate can be calibrated on the Crab response to determine an effective area in each of the energy bands. We deduce the spectrum of the residual emission per radian in figure \ref{fig:spec}.  There is a significant improvement compared to the values obtained by \citet{Purcell96} at least up to 100 keV. These results are in good agreement with an analysis using a fit of SPI data using the list of sources produced in this work \citep{Strong04}.  
 
\section{Conclusions}

It appears that the total Galactic emission is dominated by point sources emission.
The main results of the study can be summarized as follows:
\begin{itemize}
\item At 30 keV, a small fraction of the total Galactic emission ($\sim$ 14\%) is unaccounted for by detected sources. It can be caused by diffuse emission as well as unresolved point sources
\item No significant diffuse emission is detected at higher energies
\item The non-detection is not a strong constraint above 120 keV
\item Taking into account the large systematic errors, one can estimate that the residual diffuse emission should be less than 20\% of the total Galactic emission from 20 to 100 keV.
\end{itemize}

Those results are in rather good agreement with SPI results \citep{Strong04}. Some futher studies will be undertaken in the future to try to determine the flux of diffuse emission in this energy range. One of the main tasks is to lower the systematics. To do so one has to improve ISGRI angular response, take into account the source variability, and increase the background statistics at high latitudes to have a better count rate correction.  Ultimately, a combined SPI-IBIS approach should give the best constaints. Studies of the logN-logS distribution will also provide insights to the level of diffuse emission due to unresolved point sources. One should also keep in mind that a very important point lies in understanding the connection with the low energy measurements ($\le$20 keV) where a large amount of the emission is diffuse in nature. 
The non-thermal component of the Galactic ridge may be associated with quasi-thermal particles produced in SNR \citep{Dogiel02}. Their contribution to the interstellar Galactic emission should be falling in the soft $\gamma$-ray domain. The Inverse-Compton and positronium components should have much harder spectra and begin to dominate above a few hundred keV. 
A similar study is in progress to assess the level of diffuse emission seen by ISGRI between 15 and 20 keV. 

\section*{Acknowledgments}
R.T. would like to thank CNES for financial support.
This research has made use of the SIMBAD database, operated at CDS, Strasbourg, France
\bibliographystyle{aa}
\bibliography{diffus}

\end{document}